\documentclass[structabstract]{aa}  
\usepackage{graphicx}
\usepackage{txfonts}
\begin{document}
   \title{High-resolution double morphology of the most distant known radio quasar at $z$=6.12}

   \author{S.~Frey\inst{1,2}
           \and
	   L.I.~Gurvits\inst{3}
	   \and
	   Z.~Paragi\inst{2,3}
	   \and
	   K.\'E.~Gab\'anyi\inst{1,2,4}
          }

   \institute{F\"OMI Satellite Geodetic Observatory, P.O. Box 585, H-1592 Budapest, Hungary\\
              \email{frey@sgo.fomi.hu}
         \and
             MTA Research Group for Physical Geodesy and Geodynamics, P.O. Box 91, H-1521 Budapest, Hungary
         \and
             Joint Institute for VLBI in Europe, Postbus 2, 7990 AA Dwingeloo, The Netherlands\\
              \email{lgurvits@jive.nl, zparagi@jive.nl}
	 \and
             Institute of Space and Astronautical Science, JAXA, 3-1-1 Yoshinodai, Sagamihara, Kanagawa 229-8510, Japan\\
              \email{gabanyik@vsop.isas.jaxa.jp}
             }

   \date{Received Apr 24, 2008; accepted Apr 30, 2008}

 
  \abstract
   {The highest redshift quasars at $z\ga6$ receive considerable attention since they provide strong constraints on the growth of the earliest supermassive black holes. They also probe the epoch of reionisation and serve as ``lighthouses'' to illuminate the space between them and the observer. The source J1427+3312 ($z=6.12$) has recently been identified as the first and so far the only known radio-loud quasar at $z>6$.}
   {We investigated the compact radio structure of J1427+3312 on milli-arcsecond (mas) angular scales, to compare it with that of the 
second most distant radio-loud quasar J0836+0054 ($z=5.77$) and with lower-redshift radio quasars in general.}
   {We observed J1427+3312 in phase-reference mode with ten antennas of the European Very Long Baseline Interferometry (VLBI) Network (EVN) at 1.6~GHz on 11 March 2007 and at 5~GHz on 3 March 2007.}
   {The source was clearly detected at both frequencies. At 1.6~GHz, it shows a prominent double structure. The two components are separated by 28.3~mas, corresponding to a projected linear distance of $\sim160$~pc. Both components with sub-mJy flux densities appear resolved. In the position of the brightest component at 1.6~GHz, we detected mas-scale radio emission at 5~GHz as well. The radio spectrum of this feature is steep. The double structure and the separation of the components of J1427+3312 are similar to those of the young ($\la10^4$~yr) compact symmetric objects (CSOs).}  
   {}

   \keywords{techniques: interferometric --
             radio continuum: galaxies --
	     galaxies: active --
	     quasars: individual: J1427+3312 
             }

   \maketitle
%

\section{Introduction}

That the highest redshift quasars actually exist is their main interest. 
They provide significant constraints on the growth of the earliest supermassive ($\sim10^9$~$M_{\odot}$) black holes. They probe the Universe close to the
epoch of reionisation, and serve as ``lighthouses'' that illuminate the space between them and the observer. Until recently (e.g. Springel et al. \cite{spri05}), structure formation models in the $\Lambda$CDM Universe have been unable to produce massive quasars and galaxies whose the nuclei became active well before 1~Gyr following the Big Bang. Due to the small number [just over 20, Jiang et al. (\cite{jian08}) and references therein] of $z\sim6$ quasars known (of which only two are radio-loud), it is still an open question whether their physical properties are in general similar to those of lower redshift quasars. At present, we know little about quasars at $z\sim6$, especially about the radio-loud ones, simply because our sample is very small. There are arguments (e.g. Haiman et al. \cite{haim04}) in favour of a high surface density for $z>6$ radio-loud quasars.

The quasar FIRST~J1427385+331241 (J1427+3312 hereafter) was identified by McGreer et al. (\cite{mcgr06}) as the first radio-loud quasar at a redshift of $z>6$. An independent identification was made by Stern et al. (\cite{ster07}). The source is found in the NOAO Deep Wide-Field Survey (NDWFS) region. Earlier radio measurements by the Very Large Array (VLA) in the Faint Images of the Radio Sky at Twenty-centimeters (FIRST) survey (White et al. \cite{whit97}), the European Large-Area ISO Survey (ELAIS; Ciliegi et al. \cite{cili99}), and by the Westerbork Synthesis Radio Telescope (WSRT; de Vries et al. \cite{devr02}) gave 1.4-GHz flux densities of $1.7-2.1$~mJy for J1427+3312. The source has been detected in several infrared bands. The optical and near-infrared colours, the broad absorption lines, and the possible intrinsic X-ray absorption implied by the Chandra non-detection (Murray et al. \cite{murr05}) all suggest that dust extinction plays an important role in determining the appearance of this distant quasar (McGreer et al. \cite{mcgr06}). 
Becker et al. (\cite{beck00}) find that the majority of their sample of radio-emitting broad absorption line (BAL) quasars selected from the FIRST survey are compact on the $0\farcs2$ scale. Apparently the BAL quasars are not like the normal quasars seen edge-on. The existence  of a comparable number of flat- and steep-spectrum sources in the sample cannot be reconciled with simple unified models. Their preferred interpretation is that these objects are young or recently refueled quasars.

The high-resolution radio structure of the second most distant known radio-loud quasar (J0836+0054 at $z=5.77$) is characterised by a single compact but somewhat resolved component having a steep radio spectrum between 1.6 and 5~GHz. Dual-frequency Very Long Baseline Interferometry (VLBI) observations (Frey et al. \cite{frey03,frey05}) indicate that its radio emission originates from a region within $\sim40$~pc. The aim of the VLBI observations presented here was to reveal the radio structure of J1427+3312 on the linear scales of $\sim10-100$~pc. Assuming a flat cosmological model with $H_{\rm{0}}=70$~km~s$^{-1}$~Mpc$^{-1}$,  $\Omega_{\rm m}=0.3$, and $\Omega_{\Lambda}=0.7$, the redshift $z=6.12$ corresponds to 0.9~Gyr after the Big Bang ($<7$\% of the present age of the Universe). In this model, 1~mas angular separation is equivalent to a linear separation of 5.65~pc.

\section{VLBI observations and data reduction}

We observed J1427+3312 with the European VLBI Network (EVN) at 1.6~GHz on 11 March 2007 and at 5~GHz on 3 March 2007. The observed frequencies correspond to $\sim11$~GHz and $\sim36$~GHz in the rest frame of the quasar. The a-priori coordinates with nominal uncertainties of 520~mas and 390~mas in right ascension and declination, respectively, were taken from Ciliegi et al. (\cite{cili99}). At a recording rate of 1024~Mbit~s$^{-1}$, ten antennas of the EVN participated in the 7-hour observations: Effelsberg (Germany), Hartebeesthoek (South Africa), Jodrell Bank Mk2 (UK), Medicina, Noto (Italy), Toru\'n (Poland), Onsala (Sweden), Sheshan, Nanshan (P.R. China), and the phased array of the 14-element WSRT (The Netherlands). The data from the latter array were analysed as well, providing us with simultaneous measurements of the total flux density of the source at both frequencies. Eight intermediate frequency channels (IFs) were used in both left and right circular polarisation. The total bandwidth was 128~MHz in both polarisations. The correlation of the recorded VLBI data took place later at the EVN Data Processor at the Joint Institute for VLBI in Europe (JIVE), Dwingeloo, the Netherlands.

A mJy-level weak radio source, J1427+3312 was observed in phase-reference mode. This allowed us to increase the coherent integration time spent on the target source and thus to improve the sensitivity of the observations. Phase-referencing involves regularly interleaving observations between the target source and a nearby, bright, and compact reference source (e.g. Beasley \& Conway \cite{beas95}). The delay, delay rate, and phase solutions derived for the phase-reference calibrator were interpolated and applied to J1427+3312 within the target--reference cycle time of 7 minutes. The target source was observed for $\sim4.5$-minute intervals in each cycle. The total observing time on J1427+3312 was 3.2~h at both frequencies.

The phase-reference calibrator source we used was J1422+3223, a compact extragalactic radio source with $1\fdg36$ angular separation from J1427+3312. The J2000 right ascension ($\alpha = 14^{\rm{h}}22^{\rm{m}}30\fs378956$) and declination ($\delta = +32{\degr}23{\arcmin}10\farcs44012$) of the reference source in the International Celestial Reference Frame (ICRF) are available from the Very Long Baseline Array (VLBA) Calibrator Survey\footnote{{\tt http://www.vlba.nrao.edu/astro/calib/index.shtml}}. The positional uncertainty is 0.49~mas.

\begin{figure}
\centering
  \includegraphics[bb=60 120 553 720,width=75mm,angle=0,clip= ]{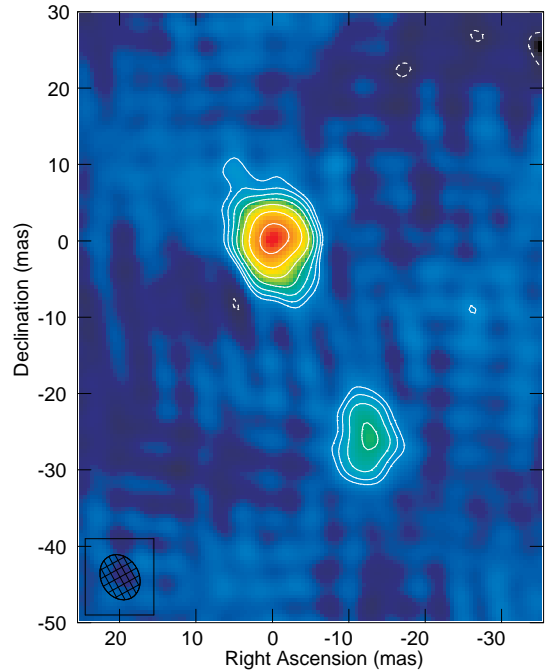}
  \caption{
The naturally weighted 1.6-GHz VLBI image of J1427+3312. The positive
contour levels increase by a factor of $\sqrt2$.
The first contours are drawn at $-50$ and 50~$\mu$Jy/beam. The peak
brightness
is 460~$\mu$Jy/beam. The Gaussian restoring beam is 6.2~mas~$\times$~5.0~mas at a major axis position angle PA=$29{\degr}$.
The coordinates are related to the brightness peak of which the absolute position is given in the text.
   }
  \label{target-1.6GHz}
\end{figure}

\begin{figure}
\centering
  \includegraphics[bb=36 162 575 678,width=75mm,angle=0,clip= ]{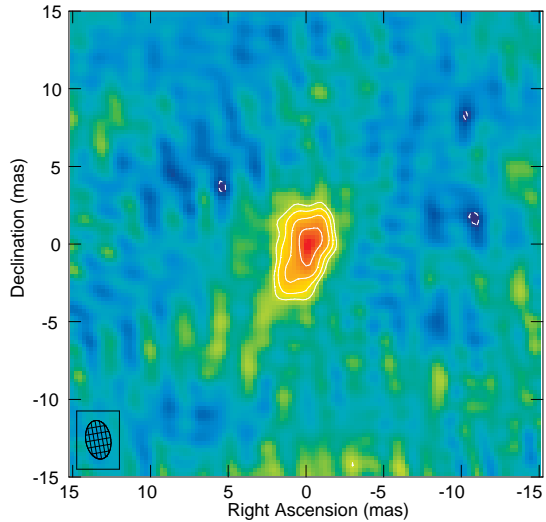}
  \caption{
The naturally weighted 5-GHz VLBI image of J1427+3312. The positive
contour levels increase by a factor of $\sqrt2$.
The first contours are drawn at $-50$ and 50~$\mu$Jy/beam. The peak
brightness
is 167~$\mu$Jy/beam. The restoring beam is 2.5~mas~$\times$~1.6~mas at PA=$10{\degr}$.
   }
  \label{target-5GHz}
\end{figure}

The US National Radio Astronomy Observatory (NRAO) Astronomical Image Processing System (AIPS, e.g. Diamond \cite{diam95}) was used for the data calibration and imaging. The visibility amplitudes were calibrated using system temperatures regularly measured at the antennas. Fringe-fitting was performed for the calibrator and fringe-finder sources (J1422+3223, J1407+2827, and J1331+3030) using 3-min solution intervals. 
The data were exported to the Caltech Difmap package (Shepherd et al. \cite{shep94}) for imaging. The conventional hybrid mapping procedure, involving several iterations of CLEANing and phase (then amplitude) self-calibration, resulted in the images and brightness distribution models for the calibrators. Overall antenna gain correction factors ($\sim15$\% or less) were determined and applied to the visibility amplitudes in AIPS.
Then fringe-fitting was repeated in AIPS, now taking the clean component models of J1422+3223 into account. The residual phase corrections that resulted from the non-pointlike structure of the phase-reference calibrator were considered this way. The solutions obtained were interpolated and applied to the target source data. The visibility data of J1427+3312, unaveraged in time and frequency, were also exported to Difmap for imaging. The naturally weighted images at 1.6~GHz (Fig.~\ref{target-1.6GHz}) and 5~GHz (Fig.~\ref{target-5GHz}) were made after several cycles of CLEANing in Difmap. The lowest contours are drawn at $\sim3\sigma$ image noise levels.
The theoretical thermal noise values were 10 and 12~$\mu$Jy/beam ($1\sigma$) at 1.6~GHz and 5~GHz, respectively.

\section{Results and discussion}

The source was clearly detected with the EVN at 1.6~GHz, showing a prominent double structure (Fig.~\ref{target-1.6GHz}). 
The two components are separated by 28.3~mas, corresponding to a projected linear distance of $\sim160$~pc. Circular Gaussian brightness distribution model fitting in Difmap indicates that both components are resolved (Table~\ref{modelfit}). The brightest one has 0.92~mJy flux density and 5.8~mas angular size (full width at half maximum, FWHM), and the weaker and more extended one has 0.62~mJy flux density and 18.6~mas angular size. In the position of the brightest component at 1.6~GHz, we detected mas-scale radio emission at 5~GHz as well (Fig.~\ref{target-5GHz}), fitted by a circular Gaussian model with 0.46~mJy flux density and 3.2 mas angular size. 
The equatorial coordinates of the brightness peak are
$\alpha_{\rm{J2000}} = 14^{\rm{h}}27^{\rm{m}}38\fs58563$ and
$\delta_{\rm{J2000}} = +33{\degr}12{\arcmin}41\farcs9252$. The positional uncertainty is determined by the phase-reference calibrator source position accuracy, the target--calibrator angular separation, the angular resolution of the interferometer array, and the signal-to-noise ratio. In our case it is estimated as better than 1~mas, due to the accurate relative astrometry with respect to the reference source position provided by the technique of phase-referencing.  

The WSRT synthesis array data that were obtained during the VLBI observations were retrieved and processed with software tools developed at JIVE for the EXPReS e-VLBI project. Comparison of the total CLEAN flux density and the WSRT flux density measurements indicates that practically the entire 
radio emission of J1427+3312 originates from the components seen in our VLBI images. Based on the fitted flux densities, the radio spectral index of the brightest feature detected at both frequencies is $\alpha=-0.6$ ($S\propto\nu^{\alpha}$, where $S$ is the flux density and $\nu$ is the frequency). J1427+3312 has also been observed and detected at 1.4~GHz with the VLBA in June 2007 (Momjian \& Carilli \cite{momj07}; Carilli \cite{cari08}).
The 1.6-GHz EVN image (Fig.~\ref{target-1.6GHz}) is fully consistent with what is seen at 1.4~GHz with the VLBA (Carilli \cite{cari08}). The significance of our dual-frequency EVN imaging results is that we can exclude the presence of a mas-scale flat-spectrum component in the source at $\sim50$~$\mu$Jy level.

\begin{table}
  \caption[]{Fitted circular Gaussian brightness distribution model parameters of J1427+3312 at 1.6~GHz {\it (top)} and 5~GHz {\it (bottom)}, and estimates of the rest-frame brightness temperatures.}
  \label{modelfit}
  \centering 
\begin{tabular}{ccccc}        
\hline\hline                 
Flux density & \multicolumn{2}{c}{Position (mas)} & Size FWHM & $T_{\rm B}$ \\
(mJy)        & North        & East                & (mas)     & ($10^6$~K) \\ 
\hline                       
0.92 &     1.37 &  $-$0.13  &   5.75 & 94 \\
0.62 & $-$24.73 & $-$11.06  &  18.62 & 6 \\
\hline
0.46 &  $-$0.29 &     0.22  &   3.22 & 15 \\
\hline   
\end{tabular}
\end{table}

The radio-loud quasar with the second highest redshift known (J0836+0054, $z=5.77$) has also been studied earlier with the EVN at 1.6 and 5~GHz (Frey et al. \cite{frey03,frey05}). It shows a single compact component. Quite remarkably, its radio spectrum is also steep ($\alpha=-0.8$). 
One could speculate that the two radio-loud $z\sim6$ quasars known to date appear to resemble the well-known gigahertz peaked-spectrum (GPS) and the somewhat larger compact steep-spectrum (CSS) sources (e.g. O'Dea \cite{odea98}). Indeed, the known parts of their radio spectra are similarly steep. A convex radio spectrum of J1427+3312 that could possibly be obtained with multi-frequency flux density measurements at lower frequencies would prove this assumption. The GPS/CSS sources are believed to be young (i.e. in the early stage of their evolution), or/and perhaps ``frustrated'' (i.e. confined by the dense interstellar medium). 
In the case of high-redshift radio galaxies, Blundell \& Rawlings (\cite{blun99}) argue that the sources we see are necessarily young ($<10^7$~yr) in a flux density-limited sample. This youth--redshift degeneracy is a consequence of the Malmquist bias (i.e. the higher the distance, the more luminous sources enter the sample) and of how the luminosity decreases with the source age.

It is not surprising that both $z\sim6$ radio quasars known to date have steep radio spectra. Early studies of the high-redshift ($z>3.5$) quasar population predicted that the space density of flat-spectrum Doppler-boosted sources declines rapidly at high redshift (Savage \& Peterson \cite{sava83}). Investigating the evolution of the quasar radio luminosity function, Dunlop \& Peacock (\cite{dunl90}) showed an abrupt (a factor of $\sim5$) cut-off of the comoving number density of the flat-spectrum radio quasars in the redshift range $2<z<4$. A similar, albeit less pronounced, trend was found for steep-spectrum sources as well. Shaver et al. (\cite{shav96}) argues for an even sharper decrease in the space density of radio-loud quasars at $z>3$. More recently Jarvis \& Rawlings (\cite{jarv00}) showed that the sample of the most luminous flat-spectrum objects is a mixture of Doppler-boosted and GPS sources. It complicates the interpretation of the redshift cut-off, which is probably more gradual beyond $z\sim2.5$.

It is interesting to note that the double structure and the separation of the components of J1427+3312 are similar to those of the young ($<10^4$~yr) compact symmetric objects (CSOs), a sub-class of GPS sources typically found in radio galaxies at much lower redshifts of $z\la1$ (e.g. Wilkinson et al. \cite{wilk94}; Polatidis et al. \cite{pola02}). If J1427+3312 is indeed an unbeamed, young symmetric object that we see in the early Universe, it might become possible to detect the expansion of its components with VLBI monitoring observations over the next decade or so. This could give us a straightforward tool for estimating the age of the radio source. If we adopt the CSO model and a typical hot spot advance speed of $\sim0.3c$ derived by Taylor et al. (\cite{tayl00}) for a sample of CSOs, we obtain a rough estimate of $\sim2000$~yr for the kinematic age of the radio source in J1427+3312.

The possibility that we see gravitationally lensed images of the same source at 1.6~GHz cannot be completely ruled out by our observations. In fact McGreer et al. (\cite{mcgr06}) found two strong intervening Mg~{\sc II} absorption systems at $z=2.18$ and $z=2.20$ in the spectrum of J1427+3312. Earlier simulation studies by e.g. Wyithe \& Loeb (\cite{wyit02}) suggest that a large fraction -- up to one third, depending on the quasar luminosity function used -- of the $z\sim6$ quasars may be magnified by gravitational lensing by a factor of 10 or more. Despite the observational efforts, no case for strong gravitational lensing is found among the known $z\sim6$ quasars so far. In the case of J1427+3312, a lens point mass of $\sim10^8$~$M_{\odot}$ located at $z=2.2$ would be needed to produce images that are separated by 28~mas. 
Strong Mg~{\sc II} absorbers are known to be associated with ``normal'' field galaxies (Steidel et al. \cite{stei94}) that could well be harbouring a central black hole of this size. However, the impact parameters are generally large, which suggests that the gravitational lensing effect of the Mg~{\sc II} systems may occur in the weak regime, enhancing only the quasar magnitudes. Using a sample of nearly 7000 strong absorbers at $0.4<z<2.2$, a recent study by M\'enard et al. (\cite{mena08}) shows that even the gravitational magnification of the background quasars due to these systems is on average negligible. We believe that strong gravitational lensing is unlikely to play a role in the case of J1427+3312.

\section{Conclusions}
We imaged J1427+3312, the highest-redshift radio-loud quasar known to date with the EVN at two frequencies (1.6 and 5~GHz). At 1.6 GHz, our phase-referenced VLBI image shows two resolved radio components with peak brightness ratio $\sim3:1$ and flux density ratio $\sim3:2$. The brightest of the two components is also detected at 5~GHz, suggesting a steep radio spectrum with spectral index $\alpha=-0.6$ between the two frequencies observed. We determined the accurate ICRF position of the source. The high-resolution radio structure of J1427+3312 is similar to that of the CSOs. This may imply a young age of the radio source, which could possibly be verified by future monitoring of the expected change in the components' separation. High-sensitivity, high-resolution radio imaging of a larger sample of $z\ga6$ radio-loud quasars still to be identified spectroscopically could provide information on how common these very compact steep-spectrum objects are in the early Universe.

\begin{acknowledgements}
The EVN is a joint facility of European, Chinese, South African, and other radio-astronomy institutes funded by their national research councils. 
This work has benefited from research funding from the European Community's sixth Framework Programme under RadioNet R113CT 2003 5058187 and from the Hungarian Scientific Research Fund (OTKA, grant
no.\ K72515). K\'EG acknowledges a fellowship received from the Japan Society for Promotion of Science. The e-VLBI developments in Europe are supported by the EC DG-INFSO 
funded Communication Network Developments project `EXPReS', Contract No.\ 02662 (http://www.expres-eu.org/).
\end{acknowledgements}

\end{document}